\documentclass[ aip, jmp, amsmath,amssymb, reprint, ]{revtex4-1}
\usepackage{graphicx}
\usepackage{dcolumn}
\usepackage{bm}
\usepackage{filecontents}
\usepackage{bbm}
\usepackage[usenames]{color}
\usepackage{afterpage}
\begin{document}

\title{Weak multiplexing in neural networks: Switching between chimera and solitary states}

\author{Maria Mikhaylenko}
\affiliation{Laboratory of Solution Chemistry of Advanced Materials and Technologies, ITMO University, 9 Lomonosova Str., Saint Petersburg 197101, Russian Federation}
\author{Lukas Ramlow}
\affiliation{ Institut f\"ur Theoretische Physik, Technische Universit\"at Berlin, Hardenbergstr. 36, 10623 Berlin, Germany}
\author{Sarika Jalan}
\affiliation{ Complex Systems Lab, Discipline of Physics, Indian Institute of Technology Indore, Khandwa Road, Simrol, Indore 453552}
\author{Anna Zakharova}
\email{anna.zakharova@tu-berlin.de}
\affiliation{ Institut f\"ur Theoretische Physik, Technische Universit\"at Berlin, Hardenbergstr. 36, 10623 Berlin, Germany}

\date{\today}

\begin{abstract}
We investigate spatio-temporal patterns occurring in a two-layer multiplex network of oscillatory FitzHugh-Nagumo neurons, where each layer is represented by a nonlocally coupled ring. We show that weak multiplexing, i.e., when the coupling between the layers is smaller than that within the layers, can have a significant impact on the dynamics of the neural network. We develop control strategies based on weak multiplexing and demonstrate how the desired state in one layer can be achieved without manipulating its parameters, but only by adjusting the other layer. We find that for coupling range mismatch weak multiplexing leads to the appearance of chimera states with different shapes of the mean velocity profile for parameter ranges where they do not exist in isolation. Moreover, we show that introducing a coupling strength mismatch between the layers can suppress chimera states with one incoherent domain (one-headed chimeras) and induce various other regimes such as in-phase synchronization or two-headed chimeras. Interestingly, small intra-layer coupling strength mismatch allows to achieve solitary states throughout the whole network.
\end{abstract}


\maketitle

\begin{quotation}
In nonlinear dynamics the paradigmatic FitzHugh-Nagumo (FHN) system is used to model the behavior of neurons. A single-layer network of nonlocally coupled oscillatory FHN neurons demonstrates various dynamic regimes including chimera states, that represent an intriguing mechanism of transition from complete coherence to complete incoherence. Here, we focus on a two-layer multiplex network and develop the tools for controlling the spatio-temporal patterns in the presence of weak coupling between the layers, i.e.\ weak multiplexing. We show that multiplexing combined with a mismatch between the layers, allows to induce chimera states in networks, where they do not appear in isolation. Our control also works in the opposite direction leading to the suppression of chimera states. Interestingly, small mismatch in the intra-layer coupling strength results in the formation of solitary states, that offer, compared to chimera states, an alternative scenario of transition from coherence to incoherence. Since multilayer structures naturally occur in neural networks (e.g., brain networks), we expect that our results can be useful for understanding and controlling of biological networks.

\end{quotation}

\section{Introduction}\label{sec:intro}

The dynamics of complex networks is one of the central issues in nonlinear dynamics \cite{SCH16,PAN15}. Coupled oscillatory units can demonstrate various types of collective behavior, including completely synchronized states, partially synchronized patterns, oscillation suppression and desynchronized dynamics. The transition from completely synchronized to completely irregular behavior can occur via different mechanisms involving special types of partial synchronization patterns. Chimera states represent an intriguing scenario, where the system spontaneously splits into coexisting domains of coherent (e.g., synchronized) and incoherent behavior, which are localized in space \cite{KUR02a,ABR04}. An alternative scenario involves solitary states where individual ``solitary'' oscillators leave the synchronous cluster at random positions in space \cite{MAI14a,JAR15}. 

Chimera states arise surprisingly in networks of completely identical units and symmetric coupling topologies \cite{ABR04,PAN15,SCH16b}. Initially detected for nonlocally coupled rings of phase oscillators, these hybrid patterns have been found for various other systems and network structures ranging from globally coupled networks \cite{YEL14,BOE15,SCH14a} and networks with power-law coupling \cite{BAN16} to irregular topologies \cite{KO08,SHA10,LAI12,YAO13,ZHU14} and hierarchical, quasi-fractal connectivities \cite{OME15,HIZ15,ULO16,TSI16,SAW17,BON18}. Solitary states have been reported for local, nonlocal and global types of coupling in networks of Kuramoto oscillators with inertia \cite{JAR15,JAR18}. 

A multilayer approach has been recently suggested to offer a better description of various real-world networks \cite{BOC14,KIV14,BIA14}. 
In multilayer networks the nodes are distributed in different layers according to the type of the relation they share. For example, in the case of a neuronal network the neurons can form different layers depending on their connectivity through a chemical link or by an ionic channel. For the investigation of the brain, the multilayer representation allows to model its structural and functional connectivity from a new viewpoint, i.e., combined with each other as layers of a multilayer network \cite{DOM17}. 

Chimera states and solitary states have been found in multilayer networks only very recently \cite{GHO16a,MAK16,AND17,BUK17,BUK18}. 
In particular, the impact of strong multiplexing, when the strength of the coupling between the layers is comparable with that inside each layer, has been investigated for coupled chaotic maps \cite{GHO16,GHO18}. In more detail, it has been shown that strong multiplexing can be used to control chimeras in networks of coupled chaotic maps \cite{GHO18}. The impact of strong multiplexing with an uncoupled layer has been investigated for a network of Hindmarsh-Rose neurons in \cite{MAJ17}, where the strong inter-layer links represent chemical connections and weak intra-layer couplings model electrical synapses. In the real-world networks it is often the case that the nodes form layers (communities or populations) where the coupling within a layer is much stronger than that across the layers. This property of community structure is common in many social and biological networks \cite{GIR02}, and in particular in neural networks \cite{HIZ16}. 

Multilayer networks not only allow for a better representation of the topology and dynamics for natural and man-made systems in comparison with isolated one-layer structures, but also open up new possibilities of control. This is especially relevant from the point of view of applications, since it is not always possible to directly manipulate a particular layer while the network it is multiplexed with may be accessible. For instance, in the case of the multilayer brain network, physical connections may be adjustable while the manipulation of the functional connectivity is much more complicated. 
Although the phenomenon of synchronization \cite{LEY17a,ZHA17,SIN15,GAM15,AND17} and formation of partial synchronization patterns \cite{KOU15,GHO16a,GHO16,JAL16,MAK16,BER17,BUK17,AND17,GHO18} have been recently considered in multilayer networks, the challenging problem of controlling chimeras by weak multiplexing, in particular, in neuronal networks has not been yet investigated. 

There occurs a question whether weak inter-layer coupling, i.e., weak multiplexing, can have a significant effect on the dynamics of the network. In particular, we address the following questions: Can weak multiplexing be used to control the spatio-temporal patterns? Does it allow to achieve desired dynamic regimes, i.e., induce, design and suppress chimera states in neural networks? Are other scenarios such as solitary states possible in the presence of weak multiplexing? What are in this situation efficient control strategies? 

In the present work we investigate a multi-layer network of coupled FitzHugh-Nagumo oscillators. We demonstrate that weak multiplexing has an essential impact on the dynamical patterns and can be used for controlling. We show that the desired states can be achieved in a particular layer when the coupling between the layers is rather weak. Different types of chimera states can be induced and suppressed. Moreover, we report the occurrence of solitary states for small intra-layer coupling strength mismatch between the layers.
Therefore, by weak multiplexing we can switch from chimera to solitary patterns. The advantage of multiplexing control we report here is that it allows to achieve the desired state in a certain layer without manipulating its parameters, and it works for weak coupling between the layers.

\section{Model}\label{sec:system}

We investigate a multiplex network consisting of two layers where each layer is represented by a nonlocally coupled ring of FitzHugh-Nagumo (FHN) oscillators. This two-dimensional system is a paradigmatic model for neural excitability \cite{MAS17}. Previously, chimera states have been found in one-layer networks consisting of coupled oscillatory \cite{OME13} and excitatory \cite{SEM16,ZAK17,ZAK17a} FHN systems. Recently weak multiplexing has been shown to play a significant role for coherence resonance in a network of coupled excitable FHN units under the influence of noise \cite{SEM18}. However, the occurrence of chimera patterns and solitary states has not yet been considered in the presence of weak multiplexing. In the present study we focus on oscillatory FHN neurons.

We consider a two-layer multiplex network, where each layer is given by a ring of $N$ nonlocally coupled FHN oscillators:\\

\begin{equation}\
\label{1}
\begin{array}{c}
\varepsilon\frac{du_{1i}}{dt}=u_{1i}-\frac{u^3_{1i}}{3}-v_{1i} +\frac{\sigma_1}{2R_1}\sum\limits_{j=i-R_1}^{i+R_1} [b_{uu}(u_{1j}-u_{1i})+\\ +b_{uv}(v_{1j}-v_{1i})] +\sigma_{12}(u_{2i}-u_{1i} ), \\
\frac{dv_{1i}}{dt}=u_{1i}+a_i+ \frac{\sigma_1}{2R_1}\sum\limits_{j=i-R_1}^{i+R_1} [b_{vu}(u_{1j}-u_{1i})+\\ +b_{vv}(v_{1j}-v_{1i})], \\
\varepsilon\frac{du_{2i}}{dt}=u_{2i}-\frac{u^3_{2i}}{3}-v_{2i} +\frac{\sigma_2}{2R_2}\sum\limits_{j=i-R_2}^{i+R_2} [b_{uu}(u_{2j}-u_{2i})+\\ +b_{uv}(v_{2j}-v_{2i})] +\sigma_{12}(u_{1i}-u_{2i} ), \\
\frac{dv_{2i}}{dt}=u_{2i}+a_i+ \frac{\sigma_2}{2R_2}\sum\limits_{j=i-R_2}^{i+R_2} [b_{vu}(u_{2j}-u_{2i})+\\ +b_{vv}(v_{2j}-v_{2i})],
\end{array}
\end{equation}

where $u_{1i}$ and $v_{1i}$ are the activator and inhibitor variables in the first(upper) layer, respectively, $i=1,...,N$ with $N$ being the total number of elements in the network. All indices are modulo $N$. 
In a similar way $u_{2i}$ and $v_{2i}$ represent the activator and inhibitor variables in the second (lower) layer, respectively. The strength of the coupling within the layer (intra-layer coupling) is given by $\sigma_1$ for the first layer and $\sigma_2$ for the second layer. The parameters $R_1$ and $R_2$ indicate the number of nearest neighbors in each direction on a ring for the first and second layer, respectively.
The coupling between the layers (inter-layer coupling) is bidirectional, diffusive and its strength is characterized by $\sigma_{12}$. Here we are mainly interested in the impact of weak multiplexing, i.e., when the inter-layer coupling $\sigma_{12}$ is smaller than the strength of the intra-layer connections $\sigma_{1}$ and $\sigma_{2}$.
We also introduce a coupling range for both layers. It is represented by the normalized number of nearest neighbours for the first (upper) layer $r_1=R_1/N$ and for the second (lower) layer $r_2=R_2/N$. A small parameter responsible for the time scale separation of fast activator and slow inhibitor is given by $\varepsilon>0$ and $a_i$ defines the excitability threshold. For an individual FHN element it determines whether the system is in the excitable ($|a_i|>1$), or oscillatory ($|a_i|<1$) regime. In the present study we assume that all elements are in the oscillatory regime ($a_{i}\equiv a=0.5$).

\begin{figure}[htbp]
\center{\includegraphics[width=0.75\linewidth, scale=0.5]{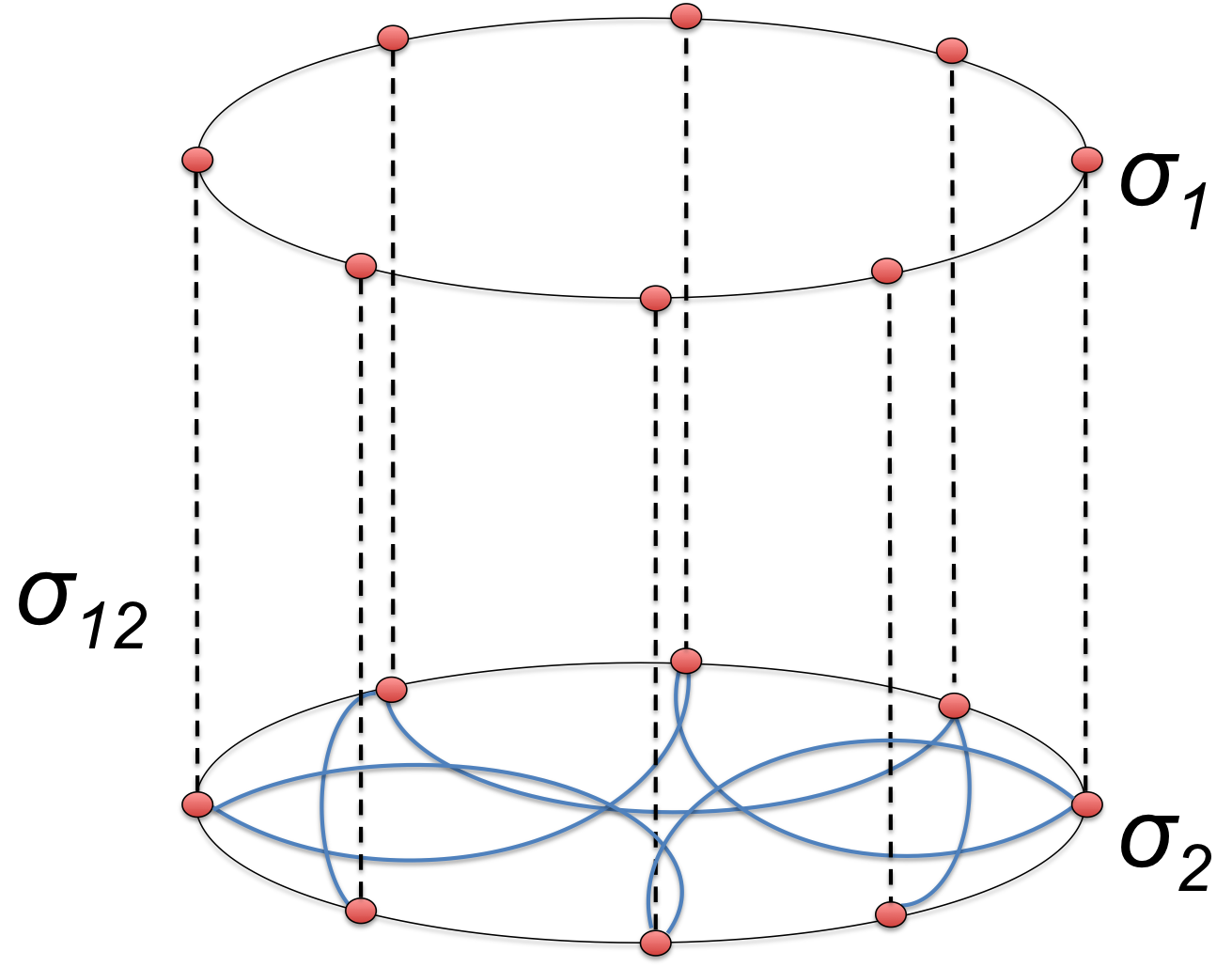}}
\caption[]{Schematic diagram showing a multiplex network consisting of two layers with different coupling range.}
\label{fig:model}  
\end{figure}

Eq.~(\ref{1}) contains not only direct, but also cross couplings between activator ($u$) and inhibitor ($v$) variables, which is modeled by a rotational coupling matrix \cite{OME13}:
\begin{equation}
B = \left(
\begin{array}{ccc}
b_{\mathrm{uu}} & & b_{\mathrm{uv}} \\
b_{\mathrm{vu}} & & b_{\mathrm{vv}}
\end{array}
\right) =
\left(
\begin{array}{ccc}
\cos \phi  & & \sin \phi \\
-\sin \phi  & & \cos \phi
\end{array}
\right),
\label{eq:Matrix_B}
\end{equation}
where $\phi\in[-\pi;\pi)$. Here we fix the parameter $\phi=\pi/2-0.1$.
Chimera states have been found for this value of $\phi$ in both the deterministic oscillatory \cite{OME13} and the noisy excitable regime \cite{SEM16,ZAK17}. Moreover, it has been shown that chimera states occurring in the excitable regime \cite{SEM16,ZAK17} are different from those detected in the oscillatory regime~\cite{OME13}. \\
 
\section{Results}

We study a multiplex network composed of two non-identical layers. In particular, we investigate the following two cases: (i) the layers are characterized by different coupling range $r_1 \neq r_2$ while the coupling strength within the layers is the same $\sigma_1=\sigma_2$; (ii) the layers have an intra-layer coupling strength mismatch, i.e.,  $\sigma_1 \neq \sigma_2$ while the coupling range is fixed and the same for both layers $r_1 = r_2$.

\subsection{Dynamics of isolated layers}
\begin{figure}[htbp]
\center{\includegraphics[width=1\linewidth]{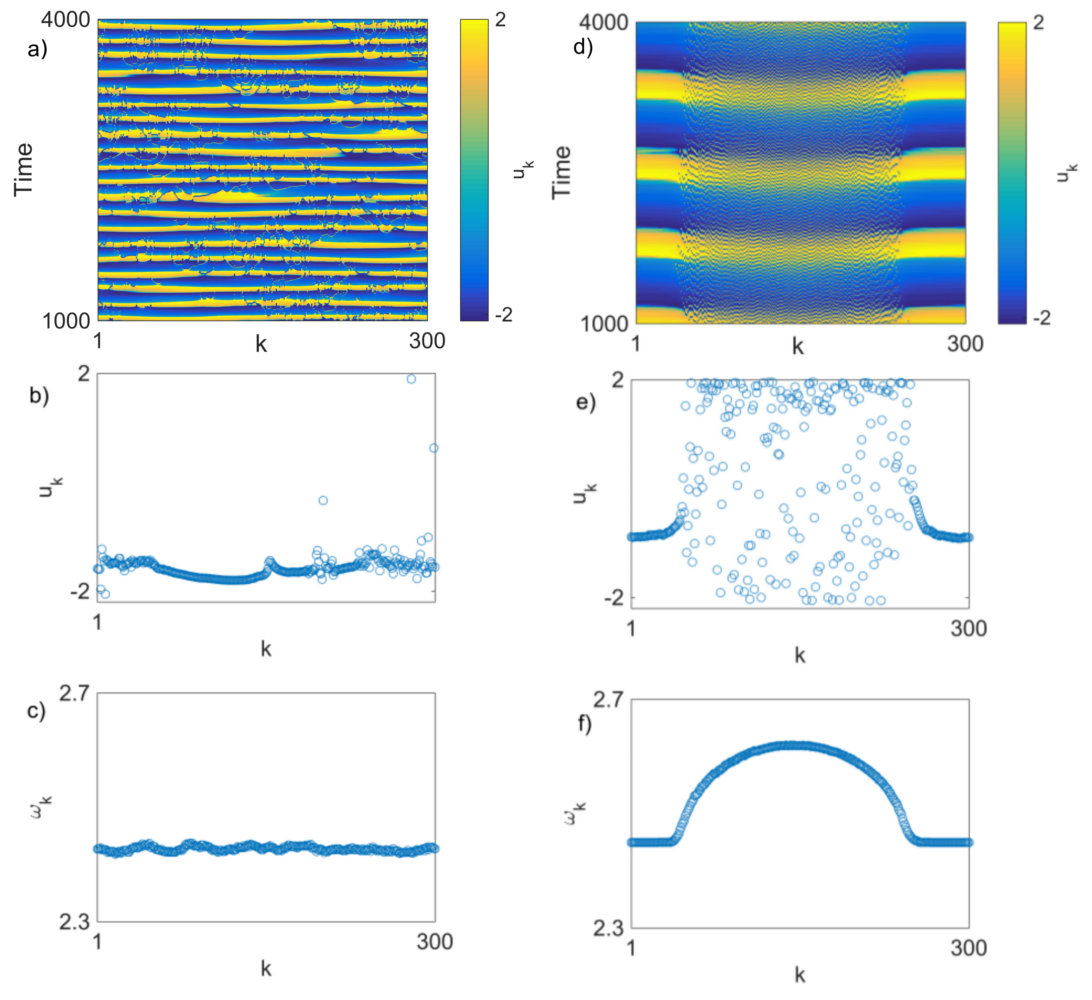}}
\caption[]{Two disconnected layers ($\sigma_{12}=0$) with different coupling range. (a), (b), (c): first (upper) layer, $r_1=0.2$; (d), (e), (f): second (lower) layer $r_2=0.35$; (a), (d): space-time plots for the variable $u_{k}$; (b), (e): snapshots for the variable $u_k$; (c), (f): mean phase velocity profiles. Initial conditions: randomly distributed on the circle $u^2 + v^2 = 4$. Other parameters: $N=300$, $\varepsilon=0.05$, $\phi=\pi/2-0.1$, $a=0.5$, $\sigma_1=\sigma_2=0.1$.}
\label{fig:no_chimera} 
\end{figure}

First, we consider the dynamics of the disconnected non-identical layers $\sigma_{12}=0$. Both of them are represented by a nonlocally coupled ring of $N=300$ identical elements and the intra-layer coupling strength is fixed $\sigma_1=\sigma_2=0.1$. We introduce the coupling range mismatch by choosing $r_1=0.2$ for the first (upper) layer and $r_2=0.35$ for the second (lower) layer. Therefore, the second layer is characterized by the higher link density compared with the first layer. The isolated layer with the smaller coupling range exhibits desynchronized dynamics (Fig.~\ref{fig:no_chimera}(a),(b),(c)). This becomes evident from the space-time plot that shows incoherent pattern (Fig.~\ref{fig:no_chimera}(a)). In the other layer chimera states are observed (Fig.~\ref{fig:no_chimera}(d),(e),(f)). Coexistence in space of well-separated synchronized and desynchronized groups of oscillating FHN neurons is clearly seen from the space-time plot (Fig.~\ref{fig:no_chimera}(d)) and the snapshot (Fig.~\ref{fig:no_chimera}(e)).  The mean phase velocity profile has a typical arc-shaped profile, a characteristic signature of chimera states (Fig.~\ref{fig:no_chimera}(f)).
The map of regimes in the ($r$, $\sigma$) parameter plane for an isolated nonlocally coupled ring of oscillatory FHN elements has been analyzed in detail in \cite{OME13}. 


\subsection{Multiplex network: coupling range mismatch.}

\begin{figure}[htbp]
\center{\includegraphics[width=0.8\linewidth]{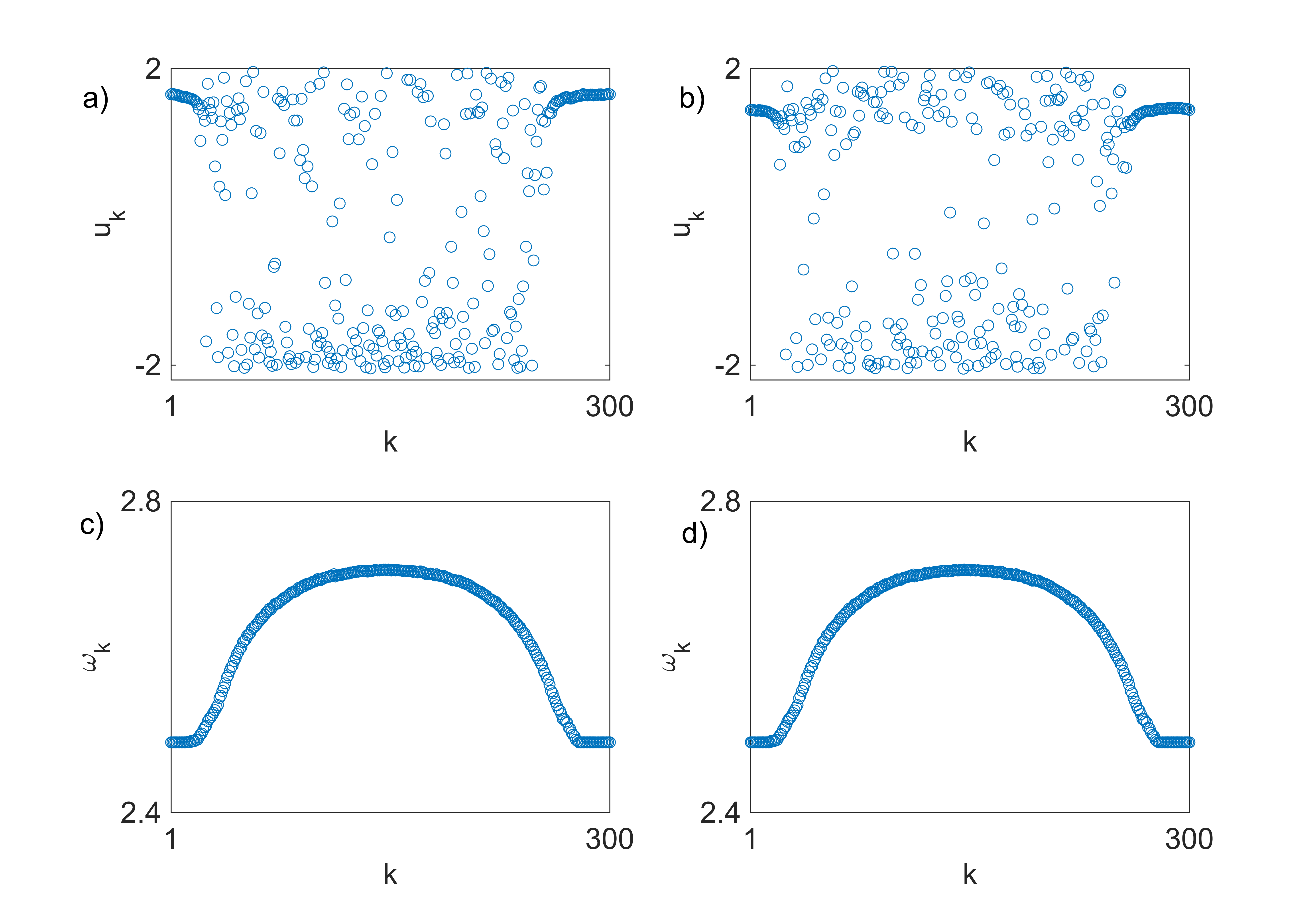}}
\caption[]{Two weakly multiplexed layers ($\sigma_{12}=0.01$) with different coupling range. (a), (c) first (upper) layer, $r_1=0.2$; (b), (d): second (lower) layer, $r_2=0.35$; (a),(b): snapshots of variable $u_k$; (c),(d): mean phase velocity profiles. Other parameters: $N=300$,  $\varepsilon=0.05$, $\phi=\pi/2-0.1$, $a=0.5$, $\sigma_1=\sigma_2=0.1$.}
\label{fig:weak_1}
\end{figure}

Next, we introduce the coupling between the non-identical layers, and investigate the impact of the inter-layer coupling strength $\sigma_{12}$ on the dynamics of the network. Interestingly, even if the inter-layer coupling $\sigma_{12}=0.01$ is much smaller than that within the layers $\sigma_{1}=\sigma_{2}=0.1$, chimera states are observed for both rings (Fig.~\ref{fig:weak_1}). One can clearly distinguish coherent and incoherent groups in the snapshots (Fig.~\ref{fig:weak_1}(a),(b)) and identify the typical arc-shape of the mean phase velocity profiles (Fig.~\ref{fig:weak_1}(c)(d)). Interestingly, the location in space of coherent and incoherent domains of the chimera pattern coincides in the two layers of the multiplex network (Fig.~\ref{fig:weak_1}).

\begin{figure}[htbp]
\begin{center}
\includegraphics[width=0.8\linewidth]{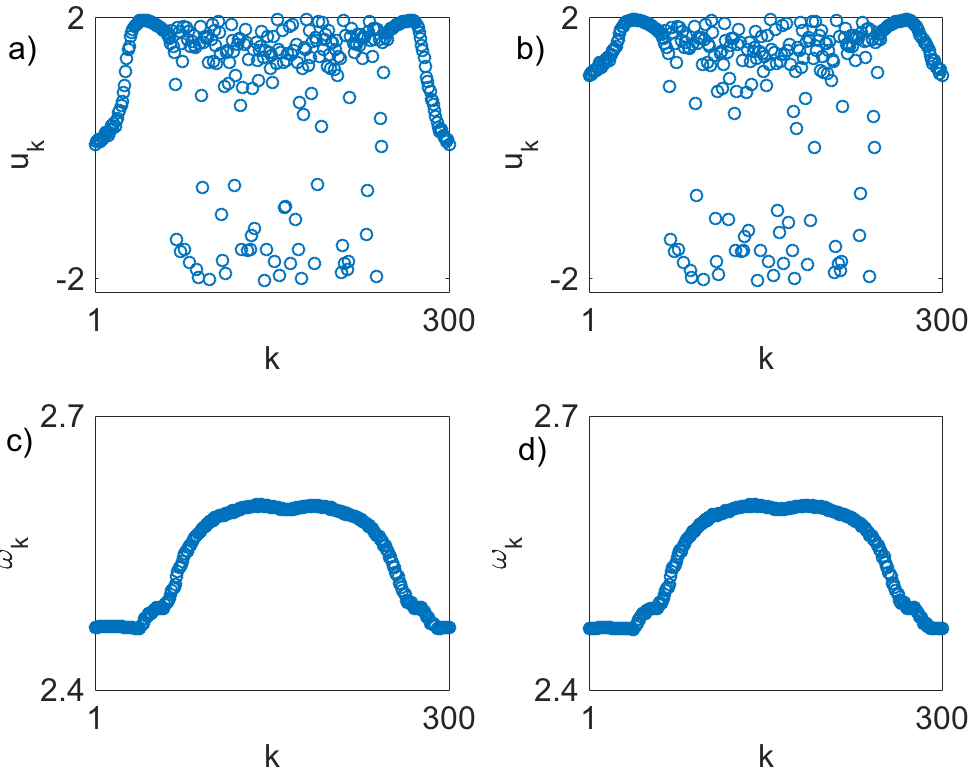}
\end{center}
\caption[]{Two strongly multiplexed layers ($\sigma_{12}=\sigma_1=\sigma_2=0.1$) with different coupling range. (a), (c): first (upper) layer, $r_1=0.2$; (b), (d): second (lower) layer, $r_2=0.35$; (a),(b): snapshots of variable $u_k$; (c),(d): mean phase velocity profiles. Other parameters: $N=300$,  $\varepsilon=0.05$, $\phi=\pi/2-0.1$, $a=0.5$.}
\label{fig:strong_1}
\end{figure}

Therefore, weak multiplexing with a denser layer allows to induce chimera states in the layer with the lower link density that does not demonstrate chimera states in isolation. Consequently, in a weakly multiplexed neural network the control of the dynamics in one of the layers can be realized without manipulating the internal parameters of its elements or the couplings between them. The control is achieved by adjusting the topology (coupling range in this particular case) of the other layer. This is relevant for the applications, since it is not always possible to directly access the desired layer while the network it is multiplexed with may be adaptable. It is important to note that the same effect can be achieved in the presence of strong multiplexing. In particular, when the strength of the coupling between the layers is equal to the inter-layer coupling $\sigma_{12}=\sigma_1=\sigma_2=0.1$, multiplexing with a denser layer induces chimera patterns in the sparser layer (Fig.~\ref{fig:strong_1}).

\begin{figure*}[t]
\center{\includegraphics[width=0.75\textwidth]{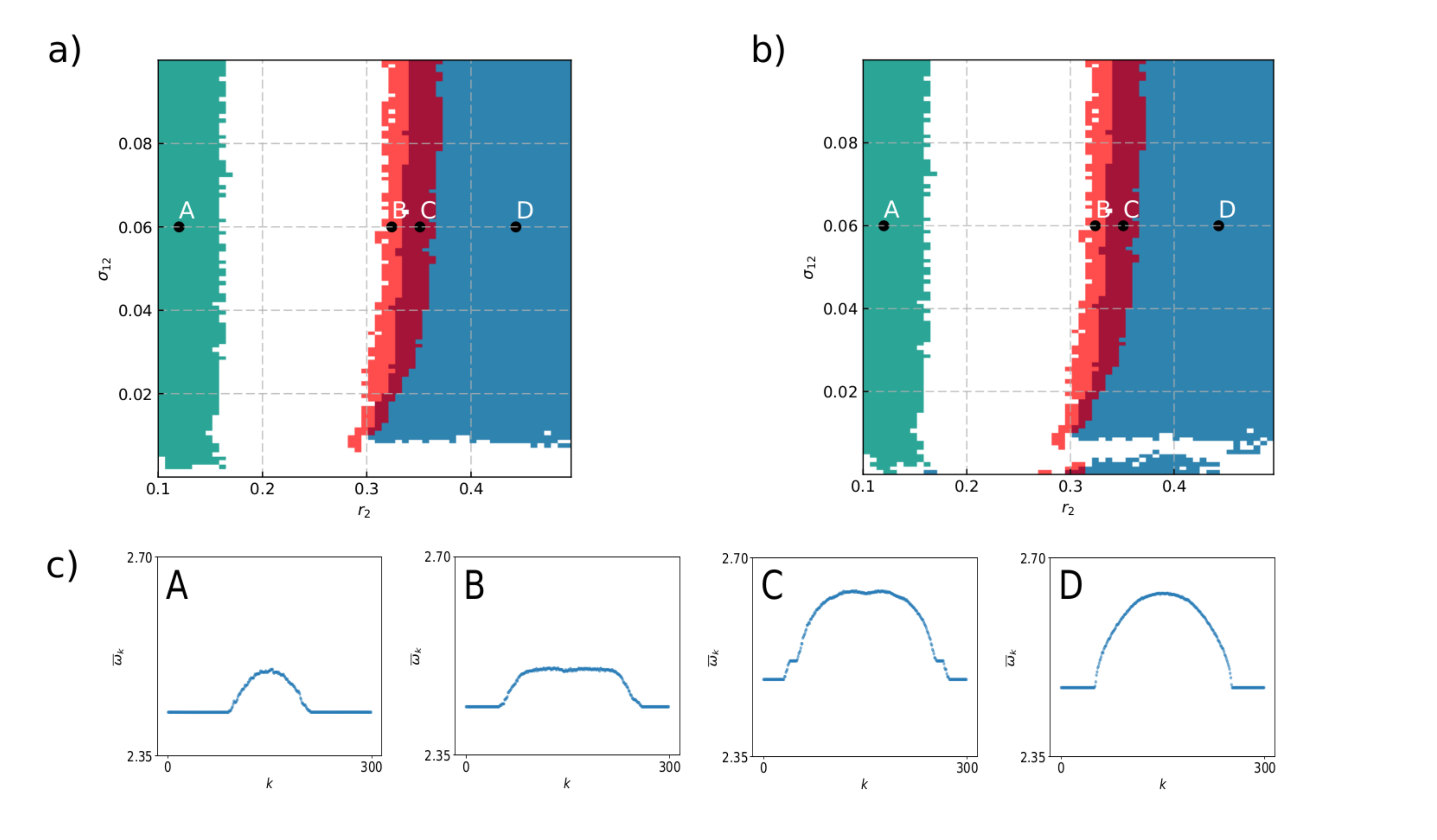}}
\caption[]{Map of regimes in the case of coupling range mismatch for (a) the first (upper) layer and (b) the second (lower) layer in the ($r_2$,$\sigma_{12}$) parameter plane. The parameters $[r_2,\sigma_{12}]$ for the selected points: A=[0.12,0.06], B=[0.323,0.06], C=[0.35,0.06], D=[0.44,0.06]. The colors stand for different chimera types; white region indicates incoherent patterns. (c): Mean phase velocity profiles for the upper layer (lower layer behaves similarly) for points A,B,C,D. Other parameters: $N=300$, $\varepsilon=0.05$, $\phi=\pi/2-0.1$, $a=0.5$, $\sigma_1=\sigma_2=0.1$,$r_1=0.2$.}
\label{fig:map_1_ab}
\end{figure*}

To get an overall view on the dynamics of the network we calculate the map of regimes in each layer for varying coupling range $r_2$ and inter-layer coupling strength $\sigma_{12}$ and keeping all other parameters fixed (Fig.~\ref{fig:map_1_ab}). In other words, we only manipulate the second (lower) layer while keeping the parameters of the first (upper) layer unchanged.

It turns out that even very weak coupling between the layers forces them to behave the same way ($\sigma_{12} > 0.007$). We detect different types of chimera states (points A,B,C,D in Fig.~\ref{fig:map_1_ab}(a),(b)) and incoherent patterns (white region in Fig.~\ref{fig:map_1_ab}(a),(b)) depending on coupling parameters.
Interestingly, chimeras can be induced in the first (upper) layer by multiplexing it with not exclusively denser layer as shown in Figs.~\ref{fig:weak_1},\ref{fig:strong_1} and Fig.~\ref{fig:map_1_ab} (see point D in Fig.~\ref{fig:map_1_ab}(a),(b) and panel D in (Fig.~\ref{fig:map_1_ab}(c)), but also with the layer characterized by lower density (point A in Fig.~\ref{fig:map_1_ab}(a),(b) and panel A in (Fig.~\ref{fig:map_1_ab}(c)). Depending on the coupling range in the second (lower) layer $r_2$, we can observe different chimera patterns. For small coupling range $r_2$ (point A in Fig.~\ref{fig:map_1_ab}(a),(b) and panel A in (Fig.~\ref{fig:map_1_ab}(c)) the chimera pattern is characterized by a smaller incoherent domain compared with those observed for larger values of $r_2$ (points B,C,D in Fig.~\ref{fig:map_1_ab}(a),(b) and panels B,C,D in (Fig.~\ref{fig:map_1_ab}(c)). Moreover, for the patterns in region A (Fig.~\ref{fig:map_1_ab}(a),(b)) the difference between the maximum frequency from the incoherent domain and that of coherent domain is smaller. The latter property also applies to the patterns in region B (Fig.~\ref{fig:map_1_ab}(a),(b)). Furthermore, the mean phase velocity of the chimeras in region B (panel B in Fig.~\ref{fig:map_1_ab}(c)) has a plateau-shaped profile while chimeras in regions A and D (Fig.~\ref{fig:map_1_ab}(a),(b)) demonstrate a classical arc-shaped profile (panels A and D in Fig.~\ref{fig:map_1_ab}(c)). 
A distinguishing feature of chimeras in region C (panel C in (Fig.~\ref{fig:map_1_ab} (c)) is a step-like structure of the mean phase velocity profile.
The formation of such a structure can be explained by the complex synchronization cascade mechanisms recently analyzed in \cite{OME18b}.
It is important to note that the coupling range where chimeras are observed in the second (lower) layer in the multiplex network is shifted towards higher values if compared with the isolated second (lower) layer (Fig.~\ref{fig:map_1_ab}(b)). Therefore, weak multiplexing induces chimeras not only in the first (upper) layer, but also in the parameter range of the second (lower) layer where no chimeras are observed in isolation.  

\subsection{Multiplex network: coupling strength mismatch.}

\begin{figure*}[htbp]
\center{\includegraphics[width=0.9\textwidth]{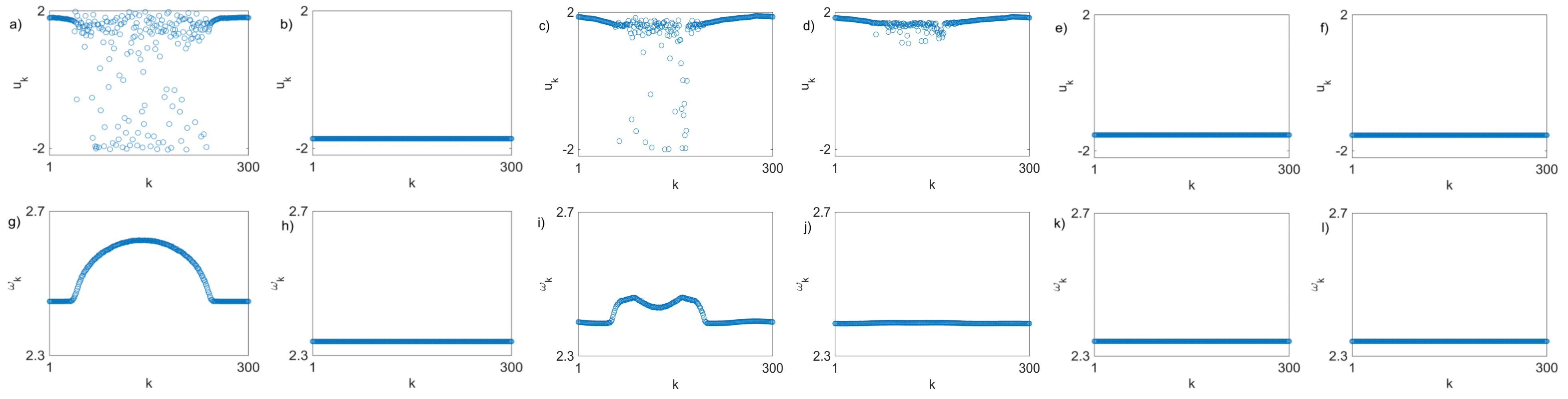}}
\caption[]{Two-layer multiplex network with intra-layer coupling strength mismatch for different values of inter-layer coupling strength $\sigma_{12}$. Snapshots of variable $u_k$ (top row) and mean phase velocity profiles (bottom row). (a),(b),(g),(h): two disconnected layers $\sigma_{12}=0$ ((a),(g) - first (upper) layer and (b),(h) - second (lower) layer); (c),(d),(i),(j): weakly coupled layers $\sigma_{12}=0.01$ ((c),(i) - first (upper) layer and (d),(j) - second (lower) layer); (e),(f),(k),(l): weakly coupled layers $\sigma_{12}=0.05$ ((e),(k) - first (upper) layer and (f),(l) - second (lower) layer). Other parameters: $N=300$, $\varepsilon=0.05$, $a=0.5$, $\phi=\pi/2-0.1$, $r_1=r_2=0.35$, $\sigma_1=0.1$, $\sigma_2=0.4$.}
\label{fig:diff_sigmas_1}
\end{figure*}

Further, we consider a different type of a multiplex network with non-identical layers. We fix the link density, i.e.\ the coupling range $r$, in both rings $r_1=r_2=0.35$ and introduce a mismatch in the intra-layer coupling strength $\sigma_1 \ne \sigma_2$. In more detail, the elements in the first (upper) layer are coupled more weakly than the nodes in the second (lower) layer $\sigma_1 < \sigma_2$. Without multiplexing $\sigma_{12}=0$, 
the first (upper) layer exhibits a chimera state (Fig.~\ref{fig:diff_sigmas_1}(a),(g)).
In contrast, the strongly coupled layer ($\sigma_2=0.4$) demonstrates coherent behavior: both the snapshot of the variable $u_k$ (Fig.~\ref{fig:diff_sigmas_1}(b)) and the flat mean phase velocity profile (Fig.~\ref{fig:diff_sigmas_1}(h)) indicate synchronization. 

Once the layers are connected, chimera states in the first (upper) layer become less pronounced if compared with the isolated case. In more detail, from the snapshots of variable $u_k$ it can be seen that the size of the incoherent domain decreases (see Fig.~\ref{fig:diff_sigmas_1}(a)) for the isolated case and (Fig.~\ref{fig:diff_sigmas_1}(c)) for the multiplexed case). In the mean phase velocity profile, the difference between the maximum frequency of incoherent domain and that of coherent domain decreases and at the same time a dip is formed in the middle of the incoherent domain (Fig.~\ref{fig:diff_sigmas_1}(i)).
The occurrence of the dip can be explained by the fact that in the isolated nonlocally coupled ring the increase of the coupling strength for the fixed coupling range leads to the formation of multichimera states. In other words, the transition from a classical chimera state with one incoherent domain to a chimera pattern with two incoherent domains occurs through the formation of the dip in the incoherent domain of the mean phase velocity profile \cite{OME13}. Here such a modification of the mean phase velocity profile is caused by weak multiplexing with the strongly coupled layer. The increase of the inter-layer coupling (within the weak multiplexing range) destroys chimera states in the first (upper) layer and induces synchronization throughout the whole network: both within and across the layers (Fig.~\ref{fig:diff_sigmas_1}(e),(f),(k),(l)). Therefore, weak multiplexing allows not only to induce chimeras (as shown in Sec.\ III B), but also to suppress them. 


\begin{figure*}[htbp]
\center{\includegraphics[width=0.75\textwidth]{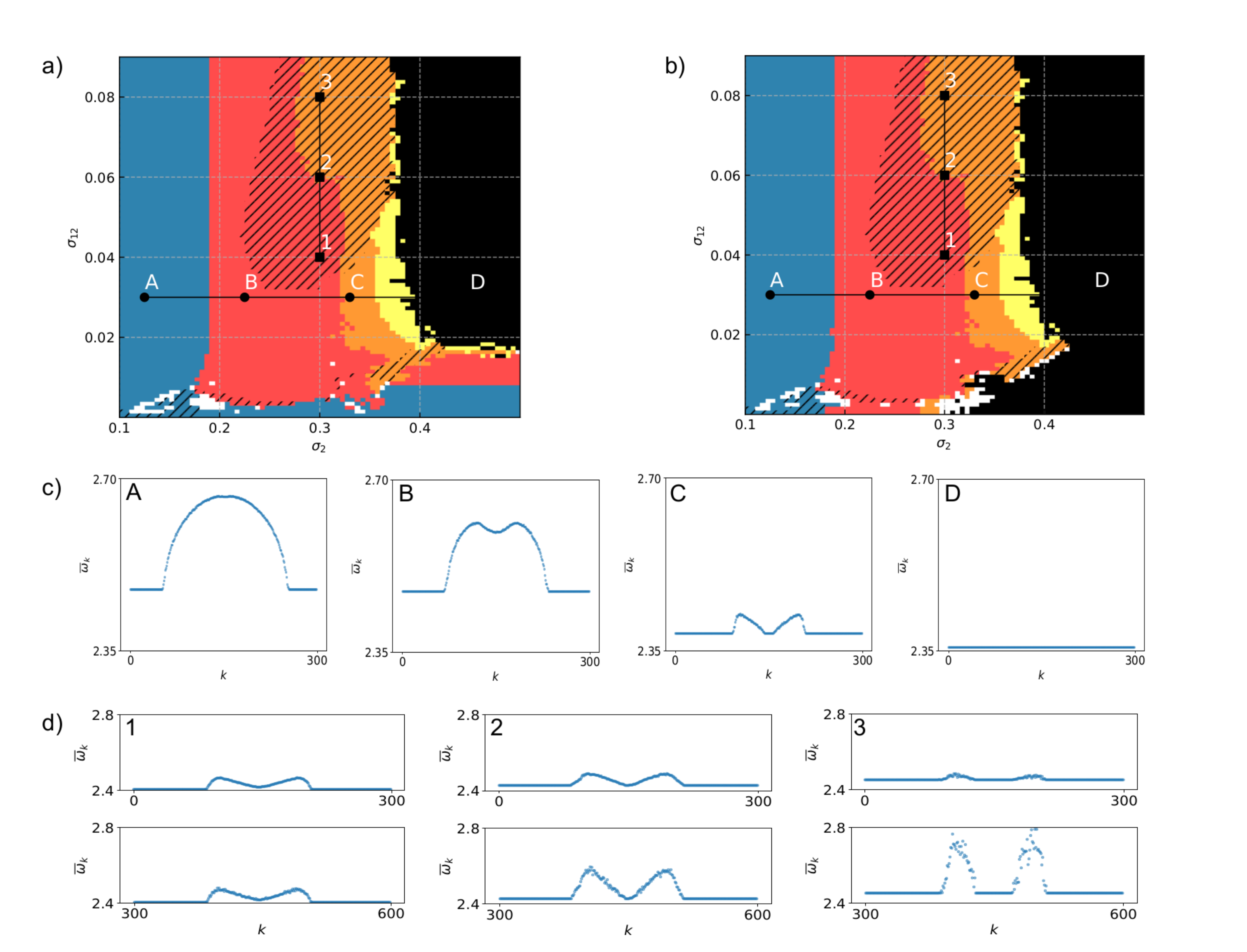}}
\caption[]{Map of regimes in the case of coupling strength mismatch for (a) the first (upper) layer and (b) the second (lower) layer
in the ($\sigma_2$,$\sigma_{12}$) parameter plane. The parameters $[\sigma_2,\sigma_{12}]$ for the selected points: A=[0.125,0.03],B=[0.225,0.03],C=[0.330,0.03],D=[0.450,0.03], 1=[0.3,0.04], 2=[0.3,0.06],3=[0.3,0.08]. Hatching marks the regions where the layers behave differently; colors stand for different chimera types; white region indicates incoherent patterns. (c): Mean phase velocity profiles for the first (upper) layer (lower layer behaves similarly) for points A,B,C,D. (d): Mean phase velocity profiles for the first (upper) layer (top row) and the second (lower) layer (bottom row) for points 1,2,3. Other parameters: $N=300$, $\varepsilon=0.05$, $\phi=\pi/2-0.1$, $a=0.5$, $\sigma_1=0.1$,$r_1=r_2=0.35$.}
\label{fig:map_2_abcd}
\end{figure*}

To get an overall view on the dynamics of the network we calculate the map of regimes in each layer for varying coupling strength $\sigma_2$ and inter-layer coupling strength $\sigma_{12}$ and keeping all other parameters fixed (Fig.~\ref{fig:map_2_abcd}). Therefore, we again manipulate only the second (lower) layer while keeping the parameters of the first (upper) layer unchanged. The difference from the Sect.~III~B is that here we adjust the strength of the intra-layer coupling and the range. In comparison with the previous case, where the coupling range was varied (see Fig.~\ref{fig:map_1_ab}), here we observe a more complex structure of regimes (Fig.~\ref{fig:map_2_abcd}). Depending on coupling parameters, the layers may behave the same way or differently (hatched region in Fig.~\ref{fig:map_2_abcd}(a),(b)). For a fixed value of the inter-layer coupling strength ($\sigma_{12}=0.03$), by changing the intra-layer coupling strength ($\sigma_2$) we observe a change in the mean phase velocity profile of chimeras (points A,B,C,D in Fig.~\ref{fig:map_2_abcd}(a),(b)). For relatively small values of $\sigma_2$ we observe classical chimeras with one incoherent domain (for example, point A in Fig.~\ref{fig:map_2_abcd}(a),(b) and panel A in Fig.~\ref{fig:map_2_abcd}(c)). The increase of $\sigma_2$ leads to the formation of a dip in the mean phase velocity profile (panel B in (Fig.~\ref{fig:map_2_abcd}(c)). When we increase $\sigma_2$ even further, the dip reaches the frequency level of the coherent domain resulting in the formation of two-headed chimera (panel C in Fig.~\ref{fig:map_2_abcd}(c)). Further increase of $\sigma_2$ leads to an in-phase synchronized regime (panel D in Fig.~\ref{fig:map_2_abcd}(c)).
Therefore, we can suppress chimeras in the first (lower) layer not only by changing the inter-layer coupling strength ($\sigma_{12}$) (as shown in Fig.~\ref{fig:diff_sigmas_1}, but also by increasing the intra-layer coupling strength in the second (lower) layer ($\sigma_2$) for a fixed value of $\sigma_{12}$ (Fig.~\ref{fig:map_2_abcd}(a)).

Compared with Sect.~III~B, where both layers show similar behavior, here the layers may behave differently depending on the coupling parameters. For the fixed value of intra-layer coupling ($\sigma_2=0.3$), the increase in inter-layer coupling strength $\sigma_{12}$ makes the layers demonstrate different mean phase velocity profiles (point 1,2,3 in Fig.~\ref{fig:map_2_abcd}(a),(b) and panels 1,2,3 in  Fig.~\ref{fig:map_2_abcd}(d)). Although, the position in space for coherent and incoherent domains in both layers stays the same and the coherent domains are synchronized, the incoherent domains start demonstrating different behavior. For example, the maximum value of mean phase velocity becomes higher in the second (upper) layer (panel 2 in  Fig.~\ref{fig:map_2_abcd}(d)). Further increase of the inter-layer coupling strength $\sigma_{12}$, when the two-headed chimera is formed, results in a better pronounced incoherent domain in the second (lower) layer, while in the first (upper) layer, the chimera is almost suppressed (panel 3 in  Fig.~\ref{fig:map_2_abcd}(d)). This result is intriguing since the two-headed chimeras are better pronounced in the second (lower) layer for the range of $\sigma_2$ values that correspond to no chimera (in-phase synchronization) for this layer in isolation. Therefore, weak multiplexing can have a dramatic effect on the dynamic regime of the network.

\subsection{Multiplex network: switching to solitary states.} 

\begin{figure}[htbp]
\center{\includegraphics[width=0.8\linewidth]{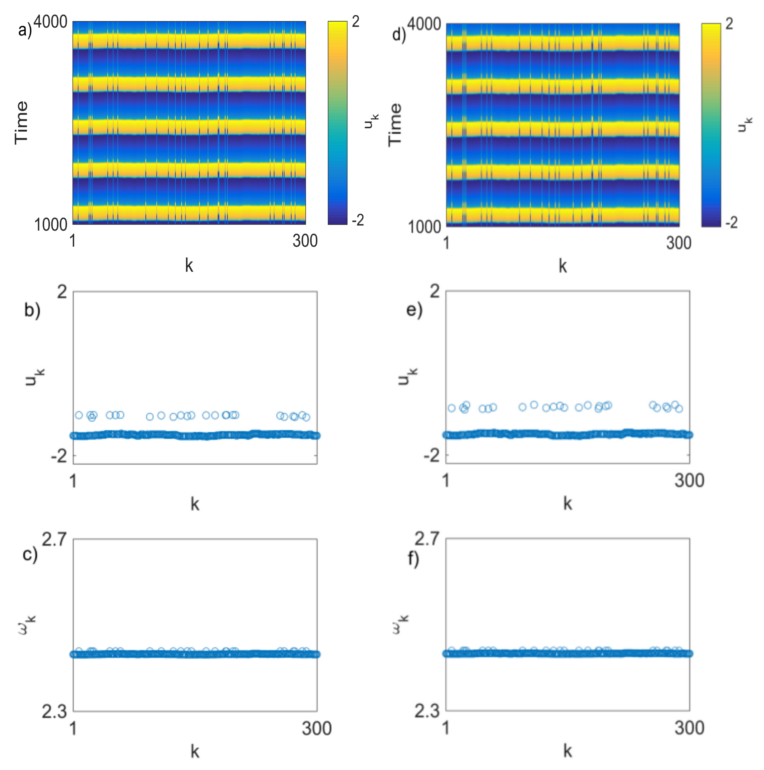}}
\caption[]{Two weakly coupled layers ($\sigma_{12}=0.05$) with small intra-layer coupling strength mismatch. (a), (b), (c): first (upper) layer, $\sigma_1=0.4$; (d), (e), (f): second (lower) layer $\sigma_2=0.3$; (a), (d): space-time plots for the variable $u_{k}$; (b), (e): snapshots for the variable $u_k$; (c), (f): mean phase velocity profiles. Other parameters: $N=300$,  $\varepsilon=0.05$, $\phi=\pi/2-0.1$, $a=0.5$, $r=0.35$.}
\label{fig:a_gamma=0,2}
\end{figure}

A solitary state is a partial synchronization pattern whose formation mechanism is different from that of a chimera state. 
In what follows we show that weak multiplexing allows not only for control of chimeras, but can also lead to the occurrence of solitary states. In particular, we demonstrate that for small intra-layer coupling strength mismatch, multiplexing induces solitary states in nonlocally coupled rings, which do not show these patterns in isolation.

In more detail, the isolated layers of nonlocally coupled FHN oscillators for the chosen set of coupling parameters exhibit complete synchronization. 
Once the two rings are coupled (and rather weak inter-layer coupling is enough) solitary states are observed in both layers (Fig.~\ref{fig:a_gamma=0,2}). We can see from the space-time plots that solitary nodes are distributed randomly along the network, a characteristic signature of solitary states (\ref{fig:a_gamma=0,2} (a)(d)). The snapshots indicate the presence of two groups of FHN elements: the coherent cluster and solitary nodes split from the synchronized group (\ref{fig:a_gamma=0,2}(b)(e)). Therefore, multiplexing allows to achieve solitary states throughout the whole network without manipulating the intra-layer coupling parameters. 

\begin{figure}[htbp]
\center{\includegraphics[width=0.6\linewidth]{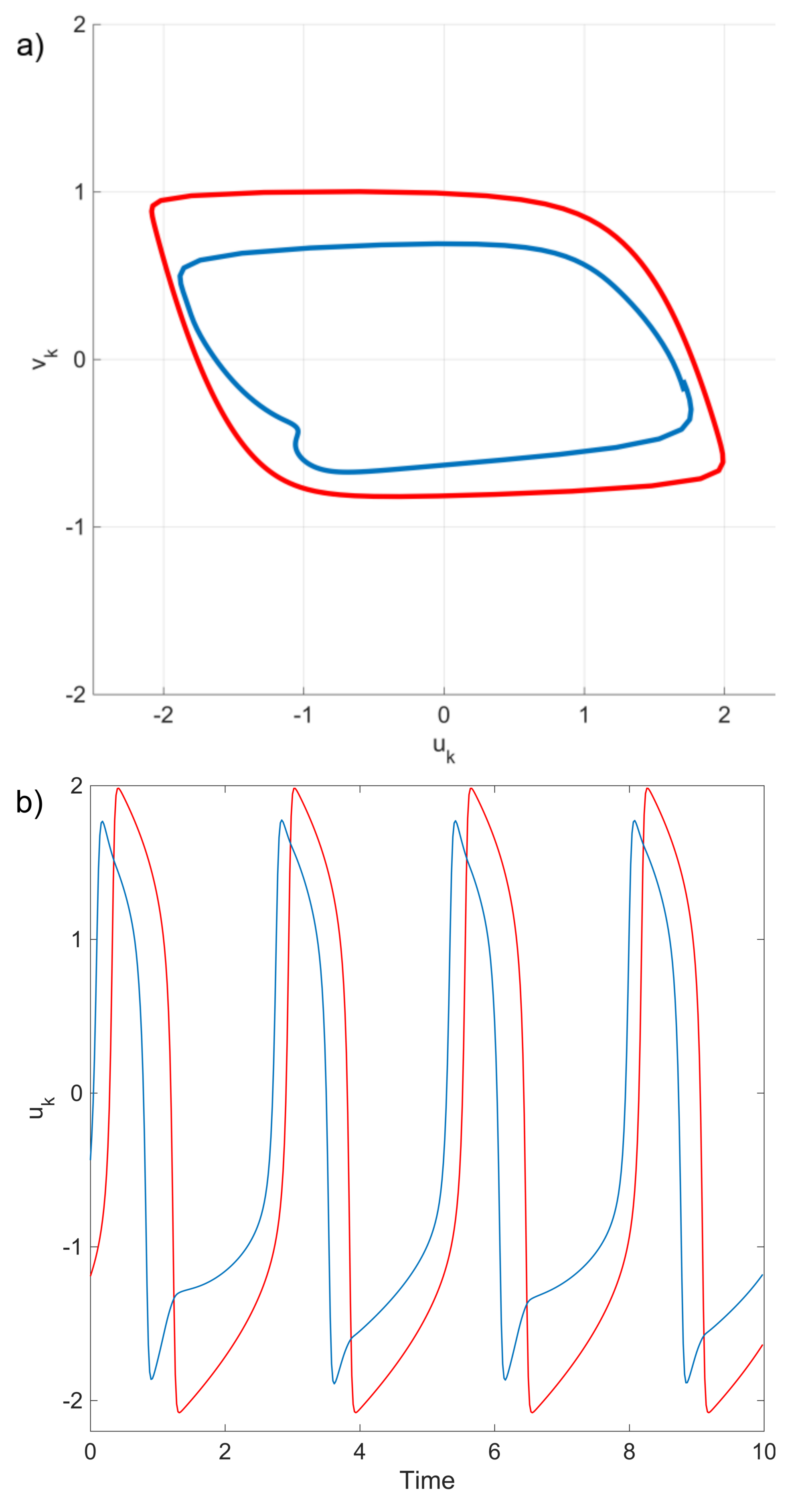}}
\caption[]{Solitary state in first (upper) layer of the multiplex network for $\sigma_{12}=0.05$. (a) phase portrait and (b) space-time plot for two selected nodes: solitary node (blu) and a node from the synchronized cluster (red). Other parameters: $N=300$, $\varepsilon=0.05$, $\phi=\pi/2-0.1$, $a=0.5$, $r=0.35$ $\sigma_1=0.4$, $\sigma_2=0.3$.}
\label{fig:NEW}
\end{figure}

It is important to point out that the solitary states we observe here for FHN neurons are different from those observed in networks of Kuramoto phase oscillators. They have a rather flat mean phase velocity profile indicating that the frequency of the solitary nodes is close to that of the synchronized cluster (Fig.~\ref{fig:a_gamma=0,2}(c),(f)). Further, the solitary nodes and the synchronized elements have different amplitudes (Fig.~\ref{fig:NEW}(a)). Another distinctive feature of the solitary patterns in FHN model is that the solitary nodes and the synchronized cluster are characterized by a small phase shift (Fig.~\ref{fig:NEW}(b)). The occurrence of the phase shift has been previously reported for globally coupled networks combining repulsive and attractive interactions \cite{MAI14a}. In contrast to that, in our case the solitary states are observed in the network where the interactions are all of the same type.

\section{Conclusions}

In conclusion, weak multiplexing can have a dramatic effect on the behavior of the network. One possible explanation for this fact can be provided by drawing an analogy with classical synchronization theory of periodic oscillations. For example, mutual synchronization of two bidirectionally coupled periodic oscillators can be achieved for a weaker connection between them if compared with external synchronization of unidirectionally coupled oscillating systems. Here we deal with two layers that are coupled mutually, therefore, weak multiplexing is enough to make them essentially influence the dynamics of each other.    

Weak multiplexing represents a powerful tool for controlling dynamic patterns in neural networks. It allows to adjust the dynamics of one layer without manipulating its parameters. In more detail, we demonstrate two control strategies: (i) by tuning the coupling range in one layer, chimeras with desired mean phase velocity profile can be induced in the other layer. Moreover, the same dynamical patterns across the layers are achieved for even very weak coupling between them. (ii) By tuning the intra-layer coupling strength we can suppress chimera states with one incoherent domain and induce a variety of other regimes, including in-phase synchronization and two-headed chimeras. Furthermore, we can make the layers behave differently. Interestingly, for small intra-layer coupling strength mismatch between the layers we can switch from a chimera to a solitary state.

In both cases the control of the dynamics for one of the layers can be realized without manipulating the internal parameters of its elements or the connections between them. The control is achieved by adjusting the coupling parameters (coupling range or coupling strength) of the other layer and varying the coupling strength between the layers within the weak multiplexing range. This is important from the point of view of applications, since it is not always possible to directly access the desired layer while the network it is multiplexed with may be adaptable. We believe that our results can be especially useful for the modeling of brain multiplex networks, where the adjustment of physical connection is feasible due to advances of modern brain surgery, while the manipulation of the functional connectivity appears to be much more complicated.

\section{Acknowledgments}
We thank Eckehard Sch\"oll, Vadim Anishchenko, Yuri Maistrenko and Iryna Omelchenko for fruitful discussions. This work was
supported by the German Academic Exchange Service (DAAD) and the Department of Science and Technology of India (DST) within the PPP project (INT/FRG/DAAD/P-06/2018). AZ acknowledges support by DFG in the framework of SFB 910.



\end{document}